\documentclass[twocolumn,epjc3]{svjour3}

\usepackage{graphicx}
\usepackage{bm}
\usepackage{amsmath}
\usepackage{amssymb}
\usepackage{hyperref}

\newcommand{\xbj}{x}
\newcommand{\zh}{z_h}
\newcommand{\nslash}{\kern 0.2 em n\kern -0.50em /}
\newcommand{\kslash}{\kern 0.2 em k\kern -0.45em /}
\newcommand{\lslash}{\kern 0.2 em l\kern -0.50em /}
\newcommand{\pslash}{\kern 0.2 em p\kern -0.50em /}
\newcommand{\Sslash}{\kern 0.2 em S\kern -0.50em /}
\newcommand{\Pslash}{\kern 0.2 em P\kern -0.50em /}
\newcommand{\Dslash}{\kern 0.2 em D\kern -0.65em /\kern 0.15em}

\newcommand{\bp}{\boldsymbol{p}_T}
\newcommand{\bP}{\boldsymbol{P}_T}
\newcommand{\bk}{\boldsymbol{k}_T}

\newcommand{\ph}{\phi_h}

\journalname{Eur. Phys. J. C}

\begin{document}

\title{On the beam spin asymmetries of electroproduction of charged hadrons off the nucleon targets}
\author{Wenjuan Mao\thanksref{addr1}
      and
      Zhun Lu\thanksref{e1,addr1}}
\thankstext{e1}{email: zhunlu@seu.edu.cn}
\institute{Department of Physics, Southeast University, Nanjing 211189, China \label{addr1}}
\maketitle

\begin{abstract}
We study the beam single-spin asymmetries $A_{LU}^{\sin\phi_h}$ for charged hadrons produced in semi-inclusive deep inelastic scattering process, by considering the $e H_1^\perp$ term and the $g^\perp D_1$ term simultaneously.
Besides the asymmetries for charged pions, for the first time we present the analysis on the asymmetries in the production of charged kaons, protons and antiprotons by longitudinally polarized leptons scattered off unpolarized proton and deuteron targets.
In our calculation we use two sets of transverse momentum dependent distributions $g^\perp(x,\bm k_T^2)$ and $e(x,\bm k_T^2)$ calculated from two different spectator models, and compare the numerical results with the preliminary data recently obtained by the HERMES Collaboration.
We also predict the beam spin asymmetries for $\pi^\pm$, $K^\pm$, $p/\bar{p}$ electroproduction in semi-inclusive deep-inelastic scattering of 12 GeV polarized electrons from unpolarized proton and deuteron targets.
\end{abstract}

\section{Introduction}
As a powerful tool to reach a more detailed understanding of the structure of hadrons, single-spin asymmetry (SSA) appearing in high energy scattering processes has attracted extensive attention in the last two decades \cite{bdr,D'Alesio:2007jt,Barone:2010ef,Boer:2011fh}.
In recent years, substantial SSAs for the electroproduction of pions and kaons in semi-inclusive deep-inelastic scattering (SIDIS) were measured by several collaborations, such as the HERMES collaboration~\cite{hermes00,hermes01,hermes03,hermes05,hermes07,hermes09,hermes10}, the Jefferson Lab (JLab)~\cite{clas04,clas10,Qian:2011py,Aghasyan:2011ha,Aghasyan:2011zz,Gohn:2014zbz} and the COMPASS collaboration~\cite{compass05,compass06,Alekseev:2008aa,compass10,Adolph:2012sn,Adolph:2012sp}.
In a particular case of SSAs, an asymmetry with a $\sin \phi_h$ modulation (the so-called beam SSA) has been observed in SIDIS by colliding the longitudinal polarized electron~\cite{clas04,Aghasyan:2011ha,Aghasyan:2011zz,Gohn:2014zbz} or positron beam~\cite{hermes07} on the unpolarized nucleon target.
Since the magnitude of the observed asymmetry with several percents cannot be explained by perturbative QCD \cite{Ahmed:1999ix}, several mechanisms have been proposed to generate such asymmetry.
One mechanism involves the $e H_1^\perp$ term~\cite{Gamberg:2003pz,Efremov:2002ut}, which indicates that the asymmetry is resulted from the coupling of the distribution $e$~\cite{Jaffe:prl67,Jaffe:npb375} with the Collins fragmentation function (FF) $H_1^\perp$~\cite{Collins:1993npb}.
Another mechanism relates to the $h_1^\perp E$ term~\cite{Yuan:2004plb}, which suggests that the beam SSA is contributed by the convolution of the Boer-Mulders function $h_1^\perp$~\cite{Boer:1998prd} and the FF $E$~\cite{Yuan:2004plb,Gamberg:2003pz}.
Apart from the above two mechanisms, a new source giving rise to the beam SSA at the twist-3 level has been found through model calculations~\cite{Afanasev:2003ze,Metz:2004epja22}.
This mechanism involves a new twist-3 transverse momentum dependent (TMD) distribution function (DF) $g^\perp$~\cite{Bacchetta:g2004plb}, which appears in the decomposition of the quark correlator if the dependence on the light-cone vector is included.
As a $T$-odd and chiral-even TMD, $g^\perp$ can be regarded as an analog of the Sivers function~\cite{Sivers:1991prd} at the twist-3 level, because both of them require quark transverse motion as well as initial- or final-state interactions~\cite{Brodsky:2002plb,Collins:2002plb,jy02} via soft-gluon exchanges to receive nonzero contributions.
Therefore, studying beam SSAs may provide a unique opportunity to unravel the role of quark spin-orbit correlation at twist 3.

In a recent work~\cite{wjmao:2012prd}, we studied the impact of $g^\perp(x,\bm k_T^2)$ on the beam SSA for neutral pion production. For this we calculated $g^\perp$ of valence quarks inside the proton using a spectator model~\cite{Bacchetta:2008prd} with scalar and axial-vector diquarks.
By comparing our results with the experimental data measured by CLAS~\cite{Aghasyan:2011ha} and HERMES~\cite{hermes07}, we found that the $T$-odd twist-3 DF $g^\perp$ may play an important role in the beam SSA in SIDIS.
In Ref.~\cite{wjmao:2013epjc}, we extended the calculations on the twist-3 TMD DFs $e$ and $g^\perp$ in the context of different spectator models for comparison.
We considered two options for the propagator of the axial-vector diquark, as well as two different relations between quark flavors and diquark types, to obtain two sets of TMD DFs.
Using the model results, we estimated the beam SSAs for neutral and charged pions at HERMES and CLAS,  by considering the $e H_1^\perp$ term and $g^\perp D_1$ term simultaneously.
Our numerical results shows that different choices for the diquark propagator will lead to different magnitudes and signs for the distribution functions, and can result in different sizes of the asymmetries.
The contributions to the beam SSAs given by the $e H_1^\perp$ term and $g^\perp D_1$ term are also quite different even in different sets.

Most recently, new preliminary measurements on the beam SSAs of charged hadrons with increased statistics were performed by the HERMES Collaboration~\cite{newhermes2013}, not only from a proton target, but also from a deuteron target.
Especially, the beam SSAs of $K^+$, $K^-$, proton and antiproton have been measured for the first time.
The new experiments adopted different kinematics from the ones in Ref.~\cite{hermes07} and extended the measurements to larger $x$ and $P_T$ regions.
The preliminary data shows that the beam SSAs for the charged pions off the proton target are slightly positive, which are consistent with our theoretical results~\cite{wjmao:2013epjc} calculated from the TMD DFs in Set 1.
For the events of charged kaons, proton and antiproton production, the data indicate that the beam SSAs are consistent with zero.
In this work, we will confront the spectator model  results\cite{wjmao:2012prd,wjmao:2013epjc} on the beam SSAs with the preliminary  data from HERMES.
Especially, we will not only present the beam SSAs for the charged pions with the new kinematic cuts at HERMES, but also give the theoretical results for the charged kaons, the proton and the antiproton, which has not been done before.
In the calculation we only consider the contribution from TMD DFs of valence quarks, therefore, the analysis on the charged kaons can be used to test the role of the sea quarks in the beam SSA.
Furthermore, we will calculate the asymmetries with both the proton and deuteron targets.
It is supposed that the contributions from the $e H_1^\perp$ term are small in the case of the deuteron target, thus the measurement with a deuteron target may provide clean evidence of the $g^\perp D_1$ term to the beam SSA, similar to the case of neutral pion production.

The rest of the paper is organized as follows.
In Section. II, we present the formalism of beam SSA in SIDIS.
In Section. III, we use two sets of TMD DFs resulted from two different spectator models to calculate the beam SSAs for charged hadrons at the new kinematic region of HERMES.
We also present the predictions on the beam SSAs in the electroproduction of  different charged hadrons at JLab with a 12 GeV electron beam.
Finally, we give our conclusion in Sec.~\ref{conclusion}.

\section{Formalism}
\begin{figure}
  \includegraphics[width=0.8\columnwidth]{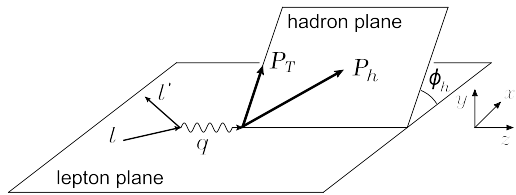}
 \caption{The kinematic configuration for the SIDIS process.
 The lepton plane ($x-z$ plane) is defined by the initial and scattered leptonic momenta, while the hadron production plane is identified by the detected hadron momentum together with the $z$ axis.}
 \label{SIDISframe}
\end{figure}

In this section, we present the formalism of beam SSA in SIDIS
\begin{align}
e^\rightarrow (\ell) \, + \, N (P) \, \rightarrow \, e' (\ell') \, + \, h (P_h) \, + \, X (P_X)\,,\label{sidis}
\end{align}
which will be applied in our phenomenological analysis later.
We adopt the reference frame where the momentum of the virtual photon defines the $z$ axis, as shown in Fig.~\ref{SIDISframe}. We use $\bk$ and $\bP$ to denote the intrinsic transverse momentum of the quark inside the nucleon and the transverse momentum of the detected hadron $h$. For he transverse momentum of the hadron with respect to the direction of the fragmenting quark, we denote it by $\bp$. Following the Trento convention~\cite{Bacchetta:2004prd}, the azimuthal angle of the hadron plane with respect to the lepton is defined as $\ph$.

The differential cross section of SIDIS for a longitudinally polarized beam with helicity $\lambda_e$ scattered off an unpolarized hadron is generally expressed as ~\cite{Bacchetta:0611265}:
\begin{align}
\label{HLT}
\frac{d\sigma}{d\xbj dy\,d\zh dP^2_T d\ph} &=\frac{2\pi \alpha^2}{\xbj y Q^2}\frac{y^2}{2(1-\varepsilon)}
 \Bigl( 1+ \frac{\gamma^2}{2\xbj} \Bigr)
  \left\{ F_{UU} \right.\nonumber\\ & \left.+ \lambda_e \sqrt{2\varepsilon(1-\varepsilon)} \sin \phi_h \,\,F^{\sin \ph}_{LU}\right\}.
\end{align}
where
$\gamma={2M x\over Q}$, and the ratio of the longitudinal and transverse photon flux $\varepsilon$ is defined as
\begin{align}
\varepsilon=\frac{1-y-\gamma^2y^2/4}{1-y+y^2/2+\gamma^2y^2/4}.
\end{align}
\begin{figure*}
  \includegraphics[width=0.9\columnwidth]{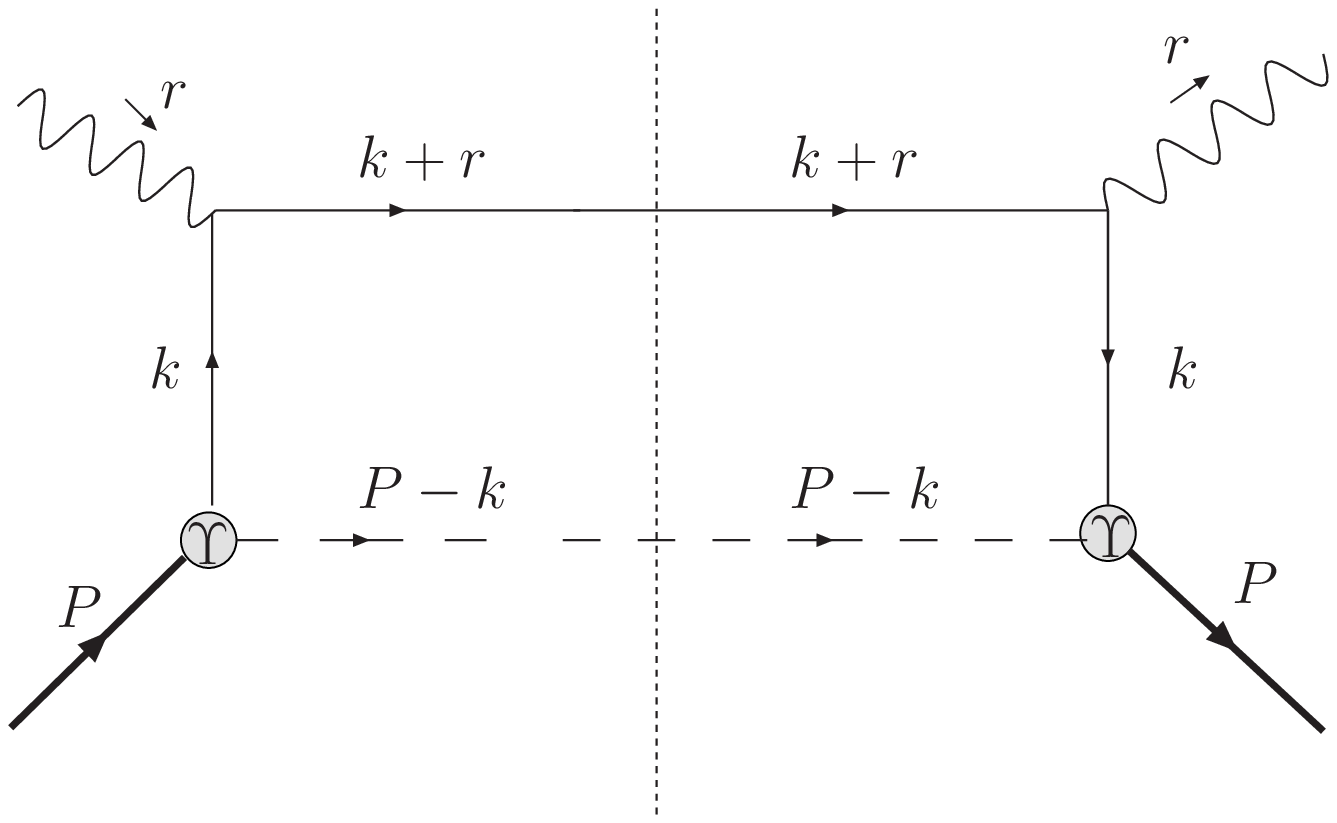}~~~~~~~~~~
  \includegraphics[width=0.9\columnwidth]{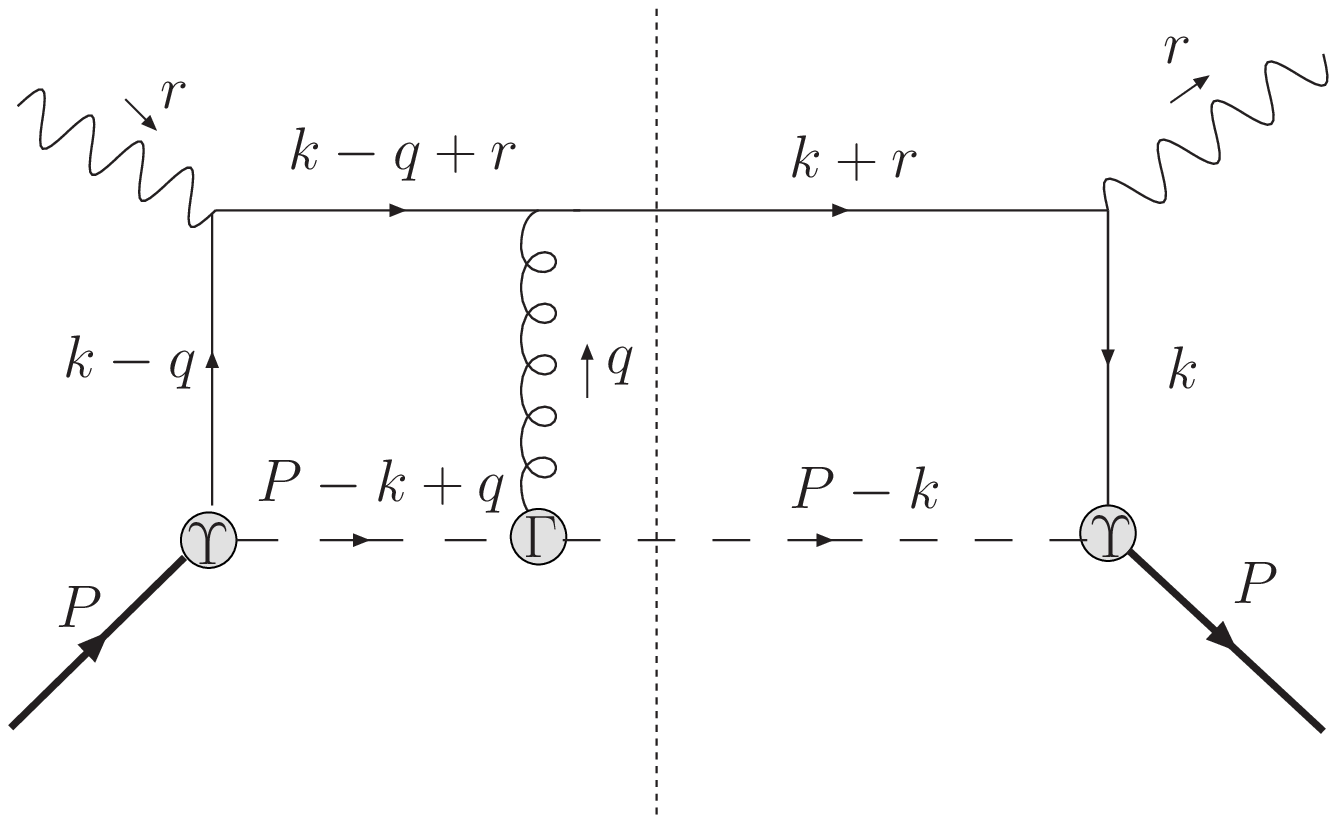}
 \caption{Diagrams in the spectator model calculation for T-even TMDs (left panel) and for T-odd TMDs (right panel). The dashed lines denote the propagators of diquarks, which can be the scalar or the axial-vector diquark.}
 \label{diagrams}
\end{figure*}

In the parton model, the unpolarized structure function $F_{UU}$ and the spin dependent structure function $F_{LU}^{\sin\phi_h}$ in Eq.~(\ref{HLT}) can be expressed as the convolutions of twist-2 and twist-3 TMD DFs and FFs, using the tree-level factorization adopted in Ref.~\cite{Bacchetta:0611265}.
With the help of the notation
\begin{align}
\mathcal{C}[w fD] &=x\sum_q e_q^2\int d^2\bm k_T\int d^2 \bm p_T\delta^2(z\bm k_T-\bm P_T+\bm p_T) \nonumber\\
&\times w(\bm k_T, \bm p_T)f^q(x,\bm k_T^2) D^q(z,\bm p_T^2),
\end{align}
we can express $F_{UU}$ and $F_{LU}^{\sin\phi_h}$ as~\cite{Bacchetta:0611265}:
\begin{align}
F_{UU} & = \mathcal{C}[f_1 D_1], \label{FUU}\\
F^{\sin \ph}_{LU} & =  \frac{2M}{Q} \,
\mathcal{C}\,
   \left[\frac{\boldsymbol{\hat{P}_{T}} \cdot \boldsymbol{p_T}}{z M_h}
         \left(\frac{M_h}{M}\,f_1\, \frac{\tilde{G^{\perp}}}{z} + \xbj\, e H_1^{\perp} \right)\right.\nonumber\\
&\left.+\frac{\boldsymbol{\hat{P}_{T}}\cdot
    \boldsymbol{k_T}}{M}\left(\frac{M_h}{M}\, h_1^{\perp} \frac{\tilde{E}}{z} +\xbj\, g^{\perp} D_1\right)\right] ,\label{FLU}
\end{align}
where $M_h$ is the mass of the final-state hadron and $\hat {\bm P}_T= {\bP\over P_T}$ with $P_T =|\bP|$.

We point out that our calculation on the structure function $F^{\sin\ph}_{LU}$ is based upon a generalization of the TMD factorization to the twist-3 level.
Therefore the correctness of our results relies on the validation of the twist-3 TMD factorization.
However the TMD factorization formalism in QCD at twist 3, or at order $1/Q$, has not been established yet.
The main challenge is that, the extension of the twist-2 factorization formula to twist 3 at high orders of $\alpha_S$ is not trivial~\cite{Gamberg:2006ru,Bacchetta:2008xw}.
Also, for the T-odd twist-3 observables, direct calculation shows that there are light-cone divergences~\cite{Gamberg:2006ru}
for which it has not been understood how to control them at order $1/Q$.
This does not necessarily mean that the twist-3 TMD factorization cannot be developed.
Further study is needed to overcome this difficulty.
Nevertheless, we will still use Eq.~(\ref{FLU}) as our starting point to study the beam SSA.

The beam SSA $A_{LU}^{\sin\phi}$ as a function of $P_T$ therefore can be written as
\begin{align}
A_{LU}^{\sin\phi_h}(P_T) &= \frac{\int dx \int dy \int dz \;\mathcal{C_F}\sqrt{2\varepsilon(1-\varepsilon)} \;F_{LU}^{\sin\phi_h}}{\int dx \int dy \int dz \;\mathcal{C_F}\;F_{UU} }, \label{asy}\\
\textrm{with}~~\mathcal{C_F}&=\frac{1}{x y Q^2}\frac{y^2}{2(1-\varepsilon)}\Bigl( 1+ \frac{\gamma^2}{2x} \Bigr).
\end{align}
The $x$-dependent and the $z$-dependent asymmetries can be defined in a similar way.

Eq.~(\ref{FLU}) shows that there are four terms giving contributions to the structure function $F^{\sin \ph}_{LU}$, which are expressed as the convolutions of the twist-3 TMD DFs or FFs with the twist-2 ones.
In the following calculation, we will neglect the $h_1^\perp \tilde{E}$ term and the $f_1 \tilde{G}^\perp$ term, based on the Wandzura-Wilczek approximation~\cite{Wandzura:1977qf}.
Thus, there are two remained terms that may give contributions to the structure function $F^{\sin \ph}_{LU}$. One is the $e H_1^\perp$ term, which has been applied to analyze the beam SSA of $\pi^+$ production in Refs.~\cite{Gamberg:2003pz,Efremov:2002ut}.
The other is the $g^\perp D_1$ term that has been adopted to calculate the beam
SSA of neutral and charged pion production~\cite{wjmao:2012prd,wjmao:2013epjc} recently.
In this work, we take both terms into consideration and finally arrive at
\begin{align}
\label{FLUa}
   F^{\sin \ph}_{LU} &\approx   \frac{2Mx}{Q} \,
  \sum_{q=u,d} e_q^2 \int d^2 \! \bk \biggl\{ \frac{\hat{\bm P}_T \cdot (\bP-z\bk)}{z M_h}\,\nonumber\\
  &\times \left[x\, e^q(x,\bk^2) H_1^{\perp q}\left(z,(\bP-z\bk)^2\right)\right]\,\nonumber\\
   &+\frac{\hat{\bm P}_T\cdot\bk} {M}
         \left[x\, g^{\perp q}(x,\bk^2) D_1^{q}\left(z,(\bP-z\bk)^2\right)\right]  \biggr\}\,.
\end{align}

For the twist-3 TMD DFs $e$ and $g^\perp$ of the $u$ and $d$ valence quarks, we apply the results from our previous work~\cite{wjmao:2013epjc}, in which we obtained two sets of TMD DFs by using two different spectator diquark models.
Among them, Set 1 is calculated from the spectator diquark model developed in Ref.~\cite{Bacchetta:2008prd}, while Set 2 is from the spectator diquark model used in Ref.~\cite{Bacchetta:plb578}.
There are two differences between these two models.
One is the choice of the propagator of the axial-vector diquark, which corresponds to the different sum of the polarization of the axial-vector diquark.
The other is the relation between quark flavors and diquark types.
In this work we will adopt both the two sets of TMD DFs to calculate beam SSAs for comparison.
The relevant diagrams for the spectator-model calculation are shown in Fig.~\ref{diagrams}, in which we denote the propagators of the diquarks by dashed lines.

In the following we explain some details on how to obtain the above mentioned two sets of TMD DFs.
In the calculation of Set 1 TMD DFs, we choose the following form for the propagator of the axial-vector diquark~\cite{Bacchetta:2008prd}
\begin{align}
 d_{\mu\nu}(P-k)  =& \,-g_{\mu\nu}\,+\, {(P-k)_\mu n_{-\nu}
 \,+ \,(P-k)_\nu n_{-\mu}\over(P-k)\cdot n_-}\,\nonumber\\
 & - \,{M_v^2 \over\left[(P-k)\cdot n_-\right]^2 }\,n_{-\mu} n_{-\nu} ,\label{d1}
\end{align}
which is the summation over the light-cone transverse polarizations of the axial-vector diquark~\cite{Brodsky:2000ii}.
At the same time, we choose the following relation between quark flavors and diquark types to obtain the TMD DFs of valence quarks:
\begin{align}
f^u=c_s^2 f^s + c_a^2 f^a,~~~~f^d=c_{a^\prime}^2 f^{a^\prime}\label{ud},
\end{align}
where $a$ and $a^\prime$ denote the
vector isoscalar diquark $a(ud)$ and the vector isovector diquark $a(uu)$, respectively, and
$c_s$, $c_a$ and $c_{a^\prime}$ are the parameters of the model.
In this calculation, the values of these model parameters are taken from Ref.~\cite{Bacchetta:2008prd}, where they were fixed by reproducing the parametrization of unpolarized~\cite{zeus} and longitudinally polarized~\cite{grsv01} parton distributions.
To calculate Set 2 TMD DFs, we adopt an alternative form for $d_{\mu\nu}$~\cite{Bacchetta:plb578}
\begin{align}
d_{\mu\nu}(P-k)  =& \,-g_{\mu\nu}\label{d2},
\end{align}
while for the relation between quark flavors and diquark types, we employ the commonly used approach in the previous spectator models calcuations~\cite{Bacchetta:plb578,Jakob:1997npa}
\begin{align}
f^u=\frac{3}{2}f^s+\frac{1}{2} f^a,~~~~f^d=f^{a^\prime}\label{set2}.
\end{align}
Here the coefficients in front of $f^X$ are obtained from the SU(4) spin-flavor symmetry of the proton wave function.
It is worthwhile to point out that another propagator of the
axial-vector diquark is investigated in~\cite{Gamberg:2007wm}, in which a complete polarization sum has been considered.

As for the Collins function $H_1^{\perp}$, we adopt the following relations for charged pions:
\begin{align}
 H_1^{\perp \pi^+/u}&=H_1^{\perp \pi^-/d}\equiv H_{1 fav}^{\perp} ,\\
 H_1^{\perp \pi^+/d}&=H_1^{\perp \pi^-/u}\equiv H_{1 unf}^{\perp} ,
\end{align}
where $H_{1 fav}^{\perp}$ and $H_{1 unf}^{\perp}$ are the favored and unfavored Collins functions, for which we apply the fitted results from Ref.~\cite{Anselmino:2008jk}.
Since currently there are no parameterized Collins functions for kaons~\cite{Bacchetta:2007wc} and proton/antiproton,
we assume that they satisfy the following relations
 \begin{align}
 \frac{H_1^{\perp K^+/u}}{D_1^{ K^+/u}} &=\frac{H_1^{\perp \pi^+/u}}{D_1^{ \pi^+/u}}, \label{kfav}\\
 \frac{H_1^{\perp p/u}}{D_1^{ p/u}} &=\frac{H_1^{\perp p/d}}{D_1^{ p/d}} =\frac{H_1^{\perp \pi^+/u}}{D_1^{ \pi^+/u}},\label{col-up}
 \end{align}
for the favored FFs and
 \begin{align}
 \frac{H_1^{\perp K^-/u}}{D_1^{ K^-/u}} &=\frac{H_1^{\perp \pi^-/u}}{D_1^{ \pi^-/u}},\label{kunfav}\\
 \frac{H_1^{\perp K^+/d}}{D_1^{ K^+/d}} &= \frac{H_1^{\perp K^-/d}}{D_1^{ K^-/d}} =  \frac{H_1^{\perp \pi^+/d}}{D_1^{ \pi^+/d}},\label{kunfav1}\\
 \frac{H_1^{\perp \bar{p}/u}}{D_1^{ \bar{p}/u}}&=\frac{H_1^{\perp \pi^-/u}}{D_1^{ \pi^-/u}},~~
 \frac{H_1^{\perp \bar{p}/d}}{D_1^{ \bar{p}/d}} = \frac{H_1^{\perp \pi^+/d}}{D_1^{ \pi^+/d}},\label{col-dp}
\end{align}
for the unfavored FFs,
which means that the ratios of favored and unfavored Collins function of the kaon and proton/antiproton are proportional to the ratios of the favored and unfavored unpolarized FFs of the pion.
For mesons, the relations in Eqs. (\ref{kfav}), (\ref{kunfav}) and (\ref{kunfav1}) may be motivated by the Artru model~\cite{Artru:1995bh}, which suggests that
all the favoured (or unfavoured) Collins function describing fragmentation into
spin-zero mesons have the same sign.
For the Collins functions of quarks fragmenting into spin-$1/2$ hadrons, currently there is no any theoretical implication or experimental constraint.
As a first approximation, we assume that they can be connected to the Collins fragmentation of mesons through Eqs.~(\ref{col-up}) and (\ref{col-dp}).
For the TMD unpolarized FF $D_1^q(z,\bp^2)$, we assume its $p_T$ dependence has a Gaussian form
\begin{align}
D_1^q\left(z,\bp^2\right)=D_1^q(z)\, \frac{1}{\pi \langle p_T^2\rangle}
\, e^{-\bm p_T^2/\langle p_T^2\rangle},
\end{align}
where $\langle p_T^2\rangle$ is the Gaussian width for $p_T^2$.
We choose $\langle p_T^2\rangle=0.2$ \textrm{GeV}$^2$ in the calculation, following the fitted result in Ref.~\cite{Anselmino:2005prd}.
For the integrated FFs $D_1^q(z)$ for different hadron production, we adopt the leading order set of the DSS parametrization~\cite{Florian:2007prd}.
\begin{figure*}
  \includegraphics[width=0.67\columnwidth]{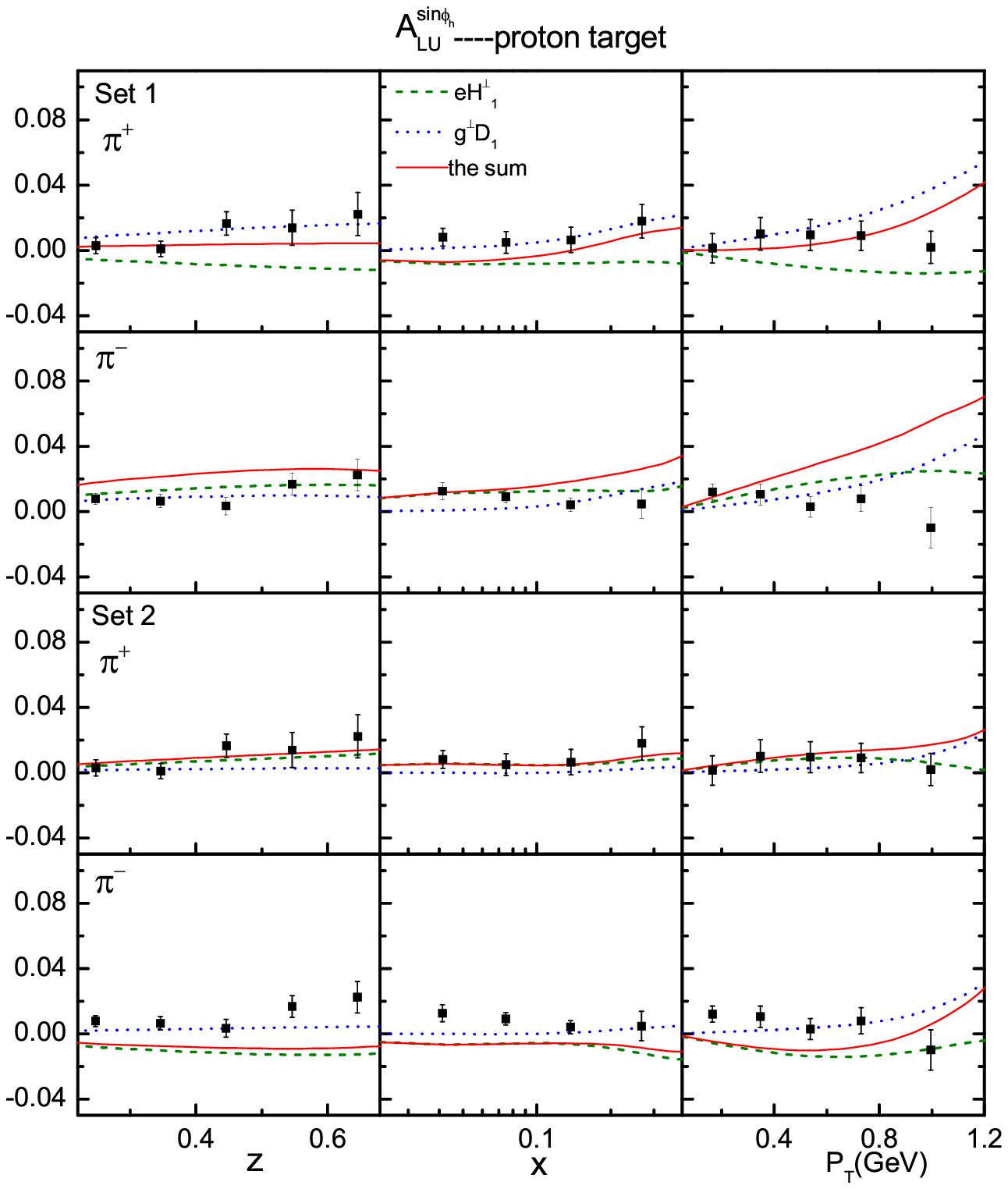}
  \includegraphics[width=0.67\columnwidth]{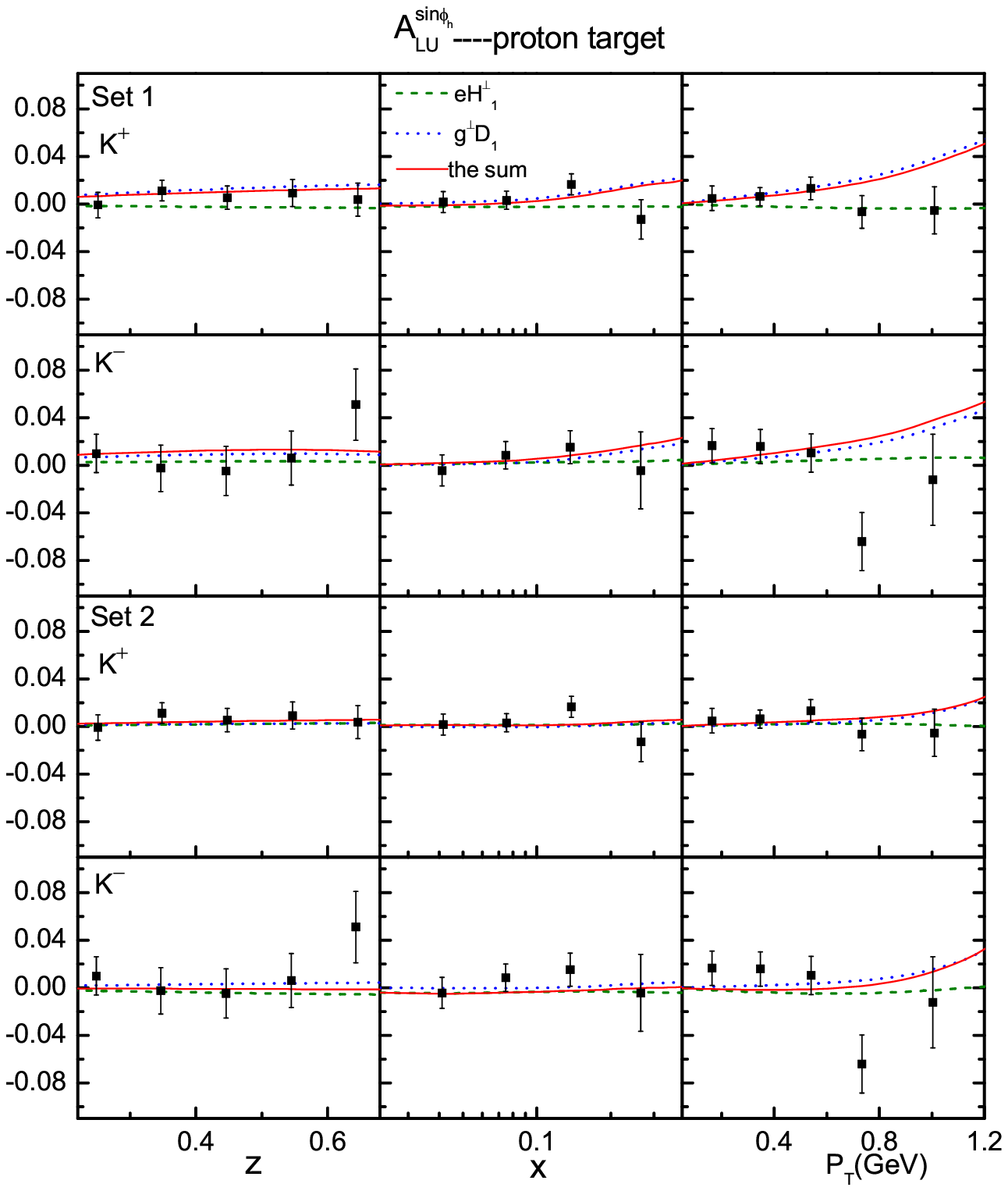}
  \includegraphics[width=0.67\columnwidth]{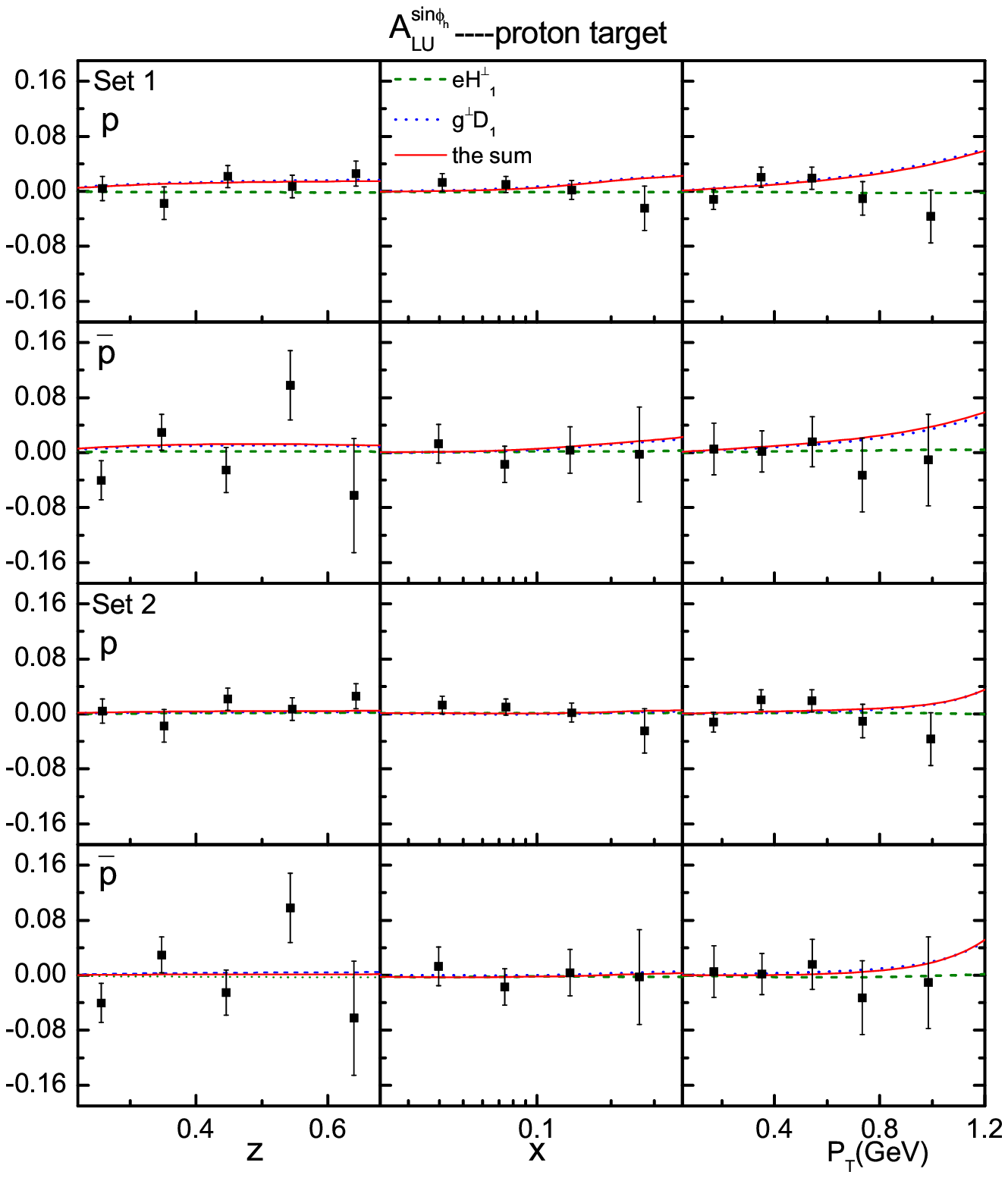}
\caption{The beam SSAs for charged pions (left panel), charged kaons (central panel) and proton/anti-proton (right panel) in SIDIS at HERMES with a proton target.
The upper panels show the results calculated from the TMD DFs in Set 1, the lower panels show the results calculated from the TMD DFs in Set 2.
The dashed, dotted and solid curves show the asymmetries from the $e H_1^\perp$ term, the $g^\perp D_1$ term and the sum of the two terms, respectively. The  preliminary data are from Ref.~\cite{newhermes2013} and the error bars include both of the systematic and statistical uncertainties.}
\label{HERMESa}
\end{figure*}

Finally, in this work, we consider the following kinematic constraints~\cite{Boglione:2011} on the intrinsic transverse momentum of the initial quarks throughout our calculation:
\begin{equation}
 \begin{cases}
k_{T}^2\leq(2-x)(1-x)Q^2, ~~~\textrm{for}~~0< x< 1 
; \\
k_{T}^2\leq \frac{x(1-x)} {(1-2x)^2}\, Q^2, ~~~~~~~~~~~~\textrm{for}~~x< 0.5.
\end{cases}\label{constraints}
 \end{equation}
They are obtained by requiring the energy of the parton to be less than the energy of the parent hadron (the first constraint) and the parton should move in the forward direction with respect to the parent hadron (the second constraint)~\cite{Boglione:2011}.
For the region $x<0.5$, there are two upper limits for $k_T^2$ at the same time; it is understood that the smaller one should be chosen.
\section{Numerical results on the beam SSAs for charged hadron production}
\label{BSAs}
\subsection{HERMES}
\begin{figure*}
  \includegraphics[width=0.67\columnwidth]{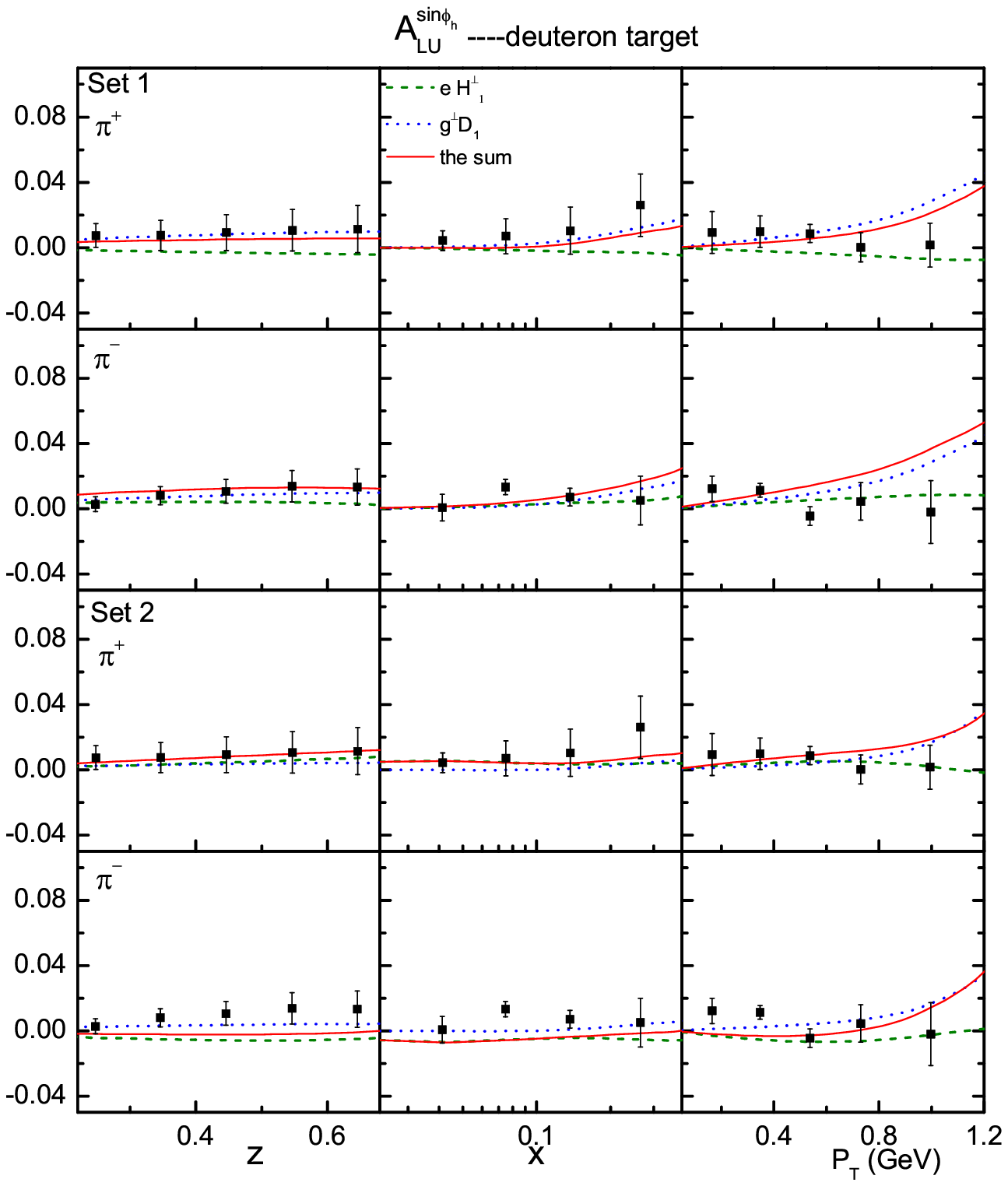}
  \includegraphics[width=0.67\columnwidth]{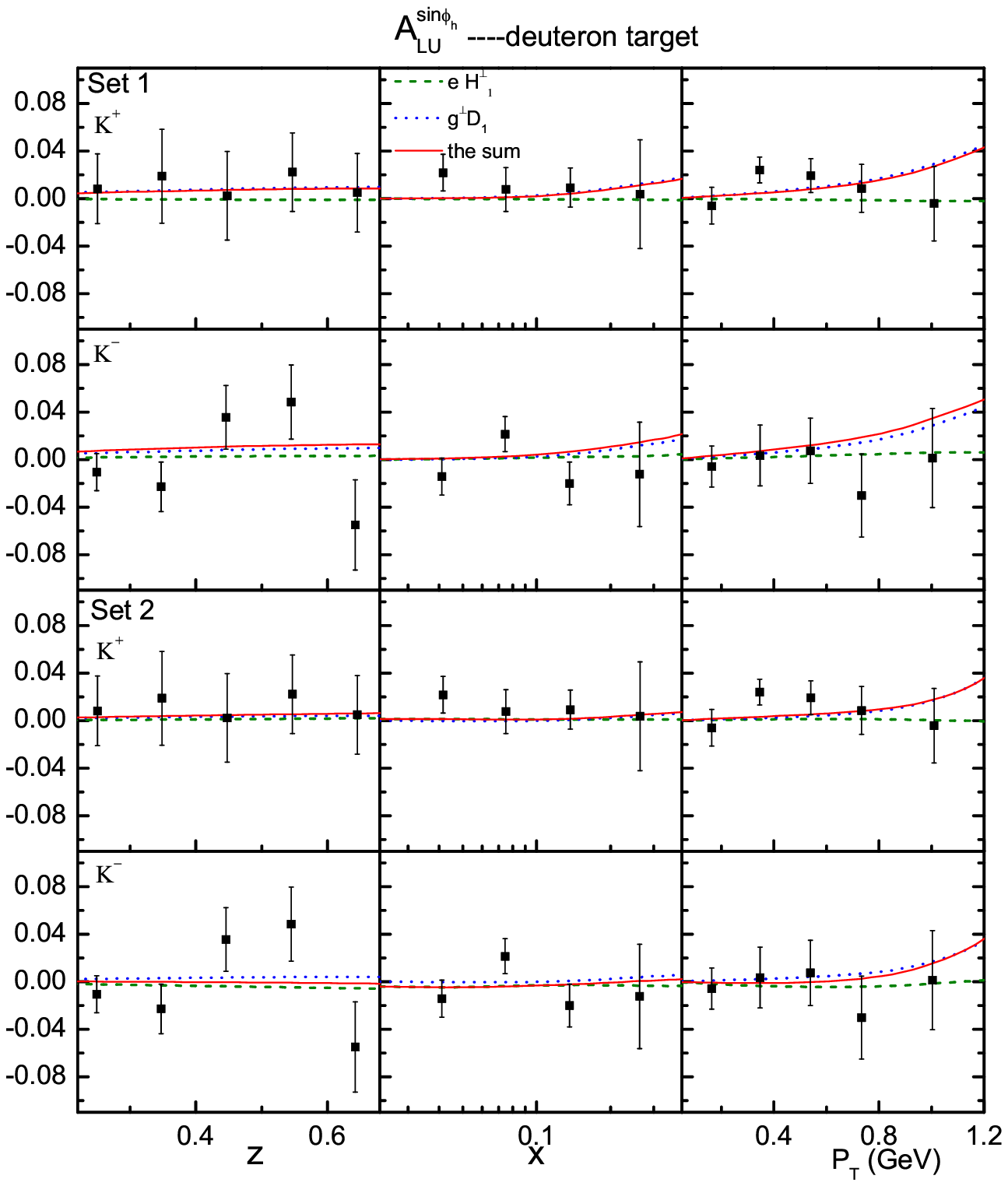}
  \includegraphics[width=0.67\columnwidth]{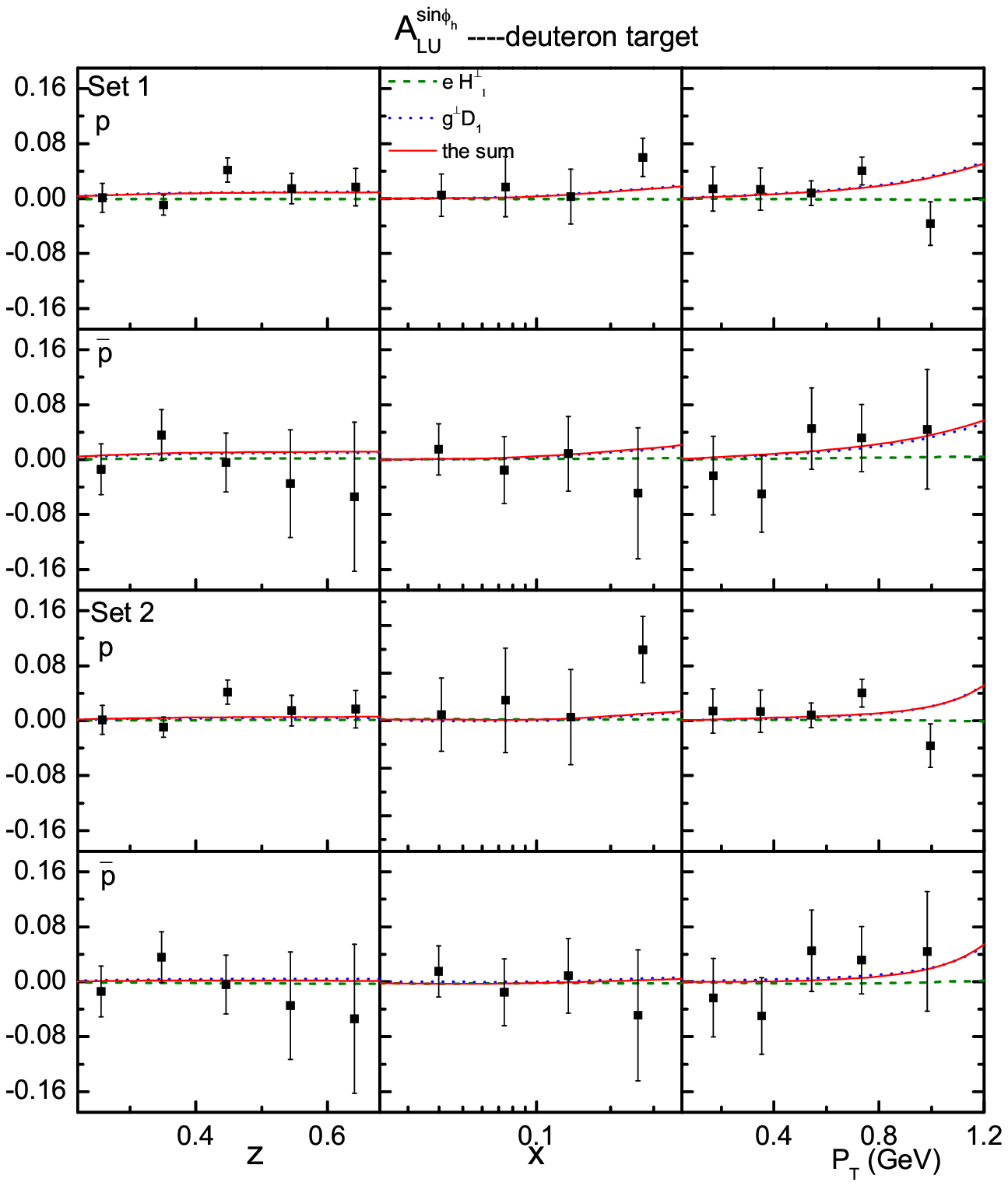}
  \caption{Similar to Fig.~\ref{HERMESa}, but with a deuteron target.}
  \label{HERMESb}
\end{figure*}
To perform numerical calculation on beam SSAs of charged hadron production in SIDIS at HERMES, we adopt the following kinematic cuts~\cite{newhermes2013}:
\begin{align}
&0.023 < x < 0.9,\,0.1 < y < 0.85, \, 0.2 < z < 0.7, \nonumber\\
& E_{\textrm{beam}}=27.6 \textrm{GeV},~~~ W^2 > 10\, \textrm{GeV}^2,, \nonumber\\
&Q^2 > 1 \textrm{GeV}^2,~~~0.05 < P_T < 1.85\,\textrm{GeV}, \nonumber\\
&\begin{cases}
2\,\textrm{GeV} < E_h < 15\, \textrm{GeV},~~~\textrm{for $\pi^{\pm}$ and $K^{\pm}$}\\
4\,\textrm{GeV} < E_h < 15\, \textrm{GeV},~~~\textrm{for $p$ and $\bar{p}$}
&\end{cases}
\end{align}
where $W$ is the invariant mass of the hadronic final states, and where $E_{\textrm{beam}}$ and $E_h$ are the energies of the electron beam and the detected final-state hadron in the target rest frame, respectively.

In the left, central, and right panels of Fig.~\ref{HERMESa}, we plot the beam SSAs for charged pions, kaons and proton/anti-proton production in SIDIS off the proton target at HERMES, as functions of $z$, $x$, and $P_T$.
The upper panels show the results calculated from the TMD DFs in Set 1, while the lower panels show the results from the TMD DFs in Set 2.
The curves are compared to the preliminary HERMES results on the asymmetries using the data collected during the years 1998-2007~\cite{newhermes2013}.
To distinguish different origins of the asymmetry, we use the dashed and dotted curves to show the contributions from the $e H_1^\perp$ term and $g^\perp D_1$ term, while the solid curves stand for the total contribution.

By comparing the theoretical results with the preliminary experimental data, we find that for $\pi^+$ production, the result in Set 2 shows a
positive asymmetry at the magnitude of $1\%$ to $2\%$, which can well describe the preliminary HERMES data.
For $\pi^-$ production, the model result from Set 1 is positive, agreeing with the sign of the preliminary HERMES data that demonstrate slightly positive asymmetry, although the calculation overestimates the data at large $x$ and large $P_T$ regions.
Our new results are the predictions on charged kaons, proton and anti-proton production, for which we obtain rather small asymmetries in both sets.
These results are consistent with the preliminary HERMES data, although the uncertainties are large.
This indicates that the valence quark approximation could be valid in the asymmetries for charged kaon, proton and anti-proton produced at HERMES.
Furthermore, the contributions from the $e H_1^\perp$ term are almost negligible in both sets.

One of the main results in this work is our prediction for the beam SSAs of charged hadrons production with a deuteron target at HERMES, as shown in Fig.~\ref{HERMESb}.
Again we plot the asymmetries for charged pions, charged kaons and proton/anti-proton production in the the left, central, and right panels.
The sizes of the asymmetries are similar to the case of the proton target. For the pion asymmetries on the deuteron target, we find that the calculation in Set 1 can well describe the preliminary data, especially for the $\pi^-$ production. Also, the agreement between the theoretical curves and the preliminary data is better than that on the proton target. Another difference from the proton target is that the dominant contributions are given by the $g^\perp D_1$ term for almost all hadrons, while the contributions from the $e H_1^\perp$ term are small compared to the $g^\perp D_1$ term. The dominance of the $g^\perp D_1$ term is more evident in Set 1.
This is not surprising because in the case of the deuteron target the $e H_1^\perp$ term contributes in the following way:
\begin{align}
\left(e^u(x,\bm k_T^2)+ e^d (x,\bm k_T^2)\right)\otimes \left(H_1^{\perp h/u} + H_1^{\perp h/d} \right),
\end{align}
where $H_1^{\perp h/u} + H_1^{\perp h/d}$ corresponds to the sum of the favored Collins function and the unfavored one.
Since the favored and the unfavored Collins functions are similar in size but opposite in sign, the $e H_1^\perp$ term contribution for the deuteron target is largely suppressed.
In the case of the charged hadron production, it would be more ideal to probe the distribution $g^\perp$ using the deuteron target than the proton target at HERMES.
\subsection{CLAS 12GeV}
\label{prediction}
\begin{figure*}
  \includegraphics[width=0.67\columnwidth]{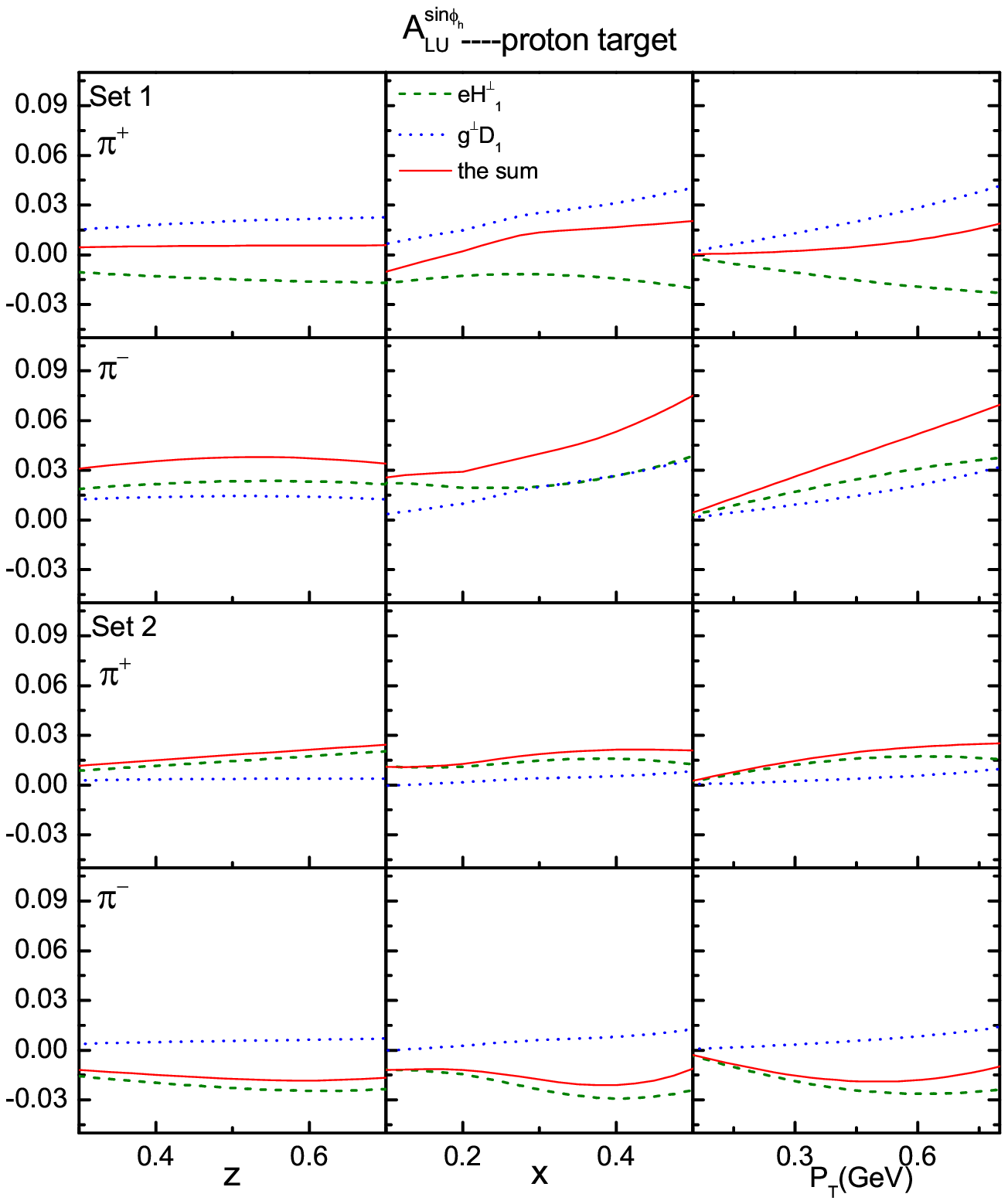}
  \includegraphics[width=0.67\columnwidth]{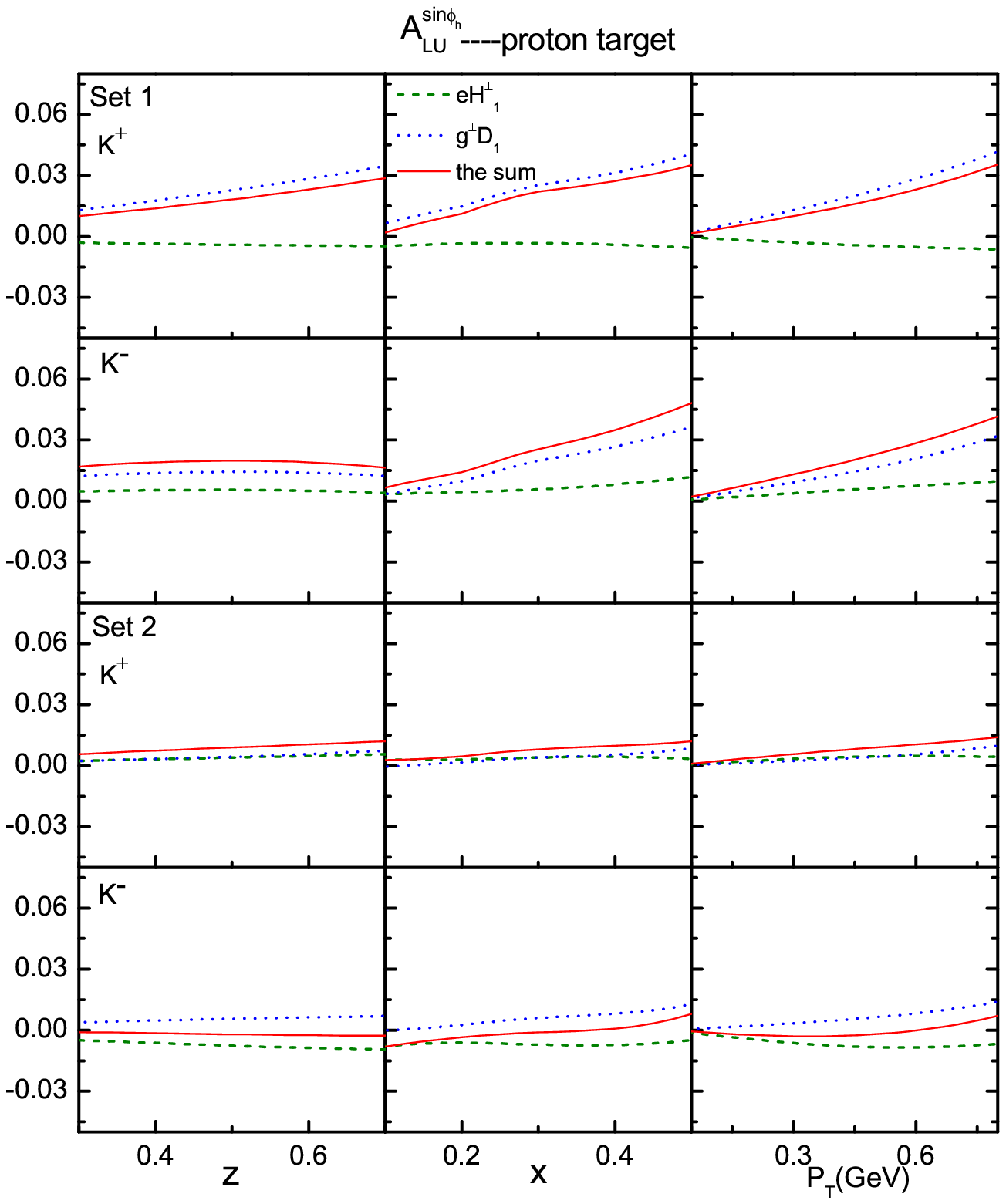}
  \includegraphics[width=0.67\columnwidth]{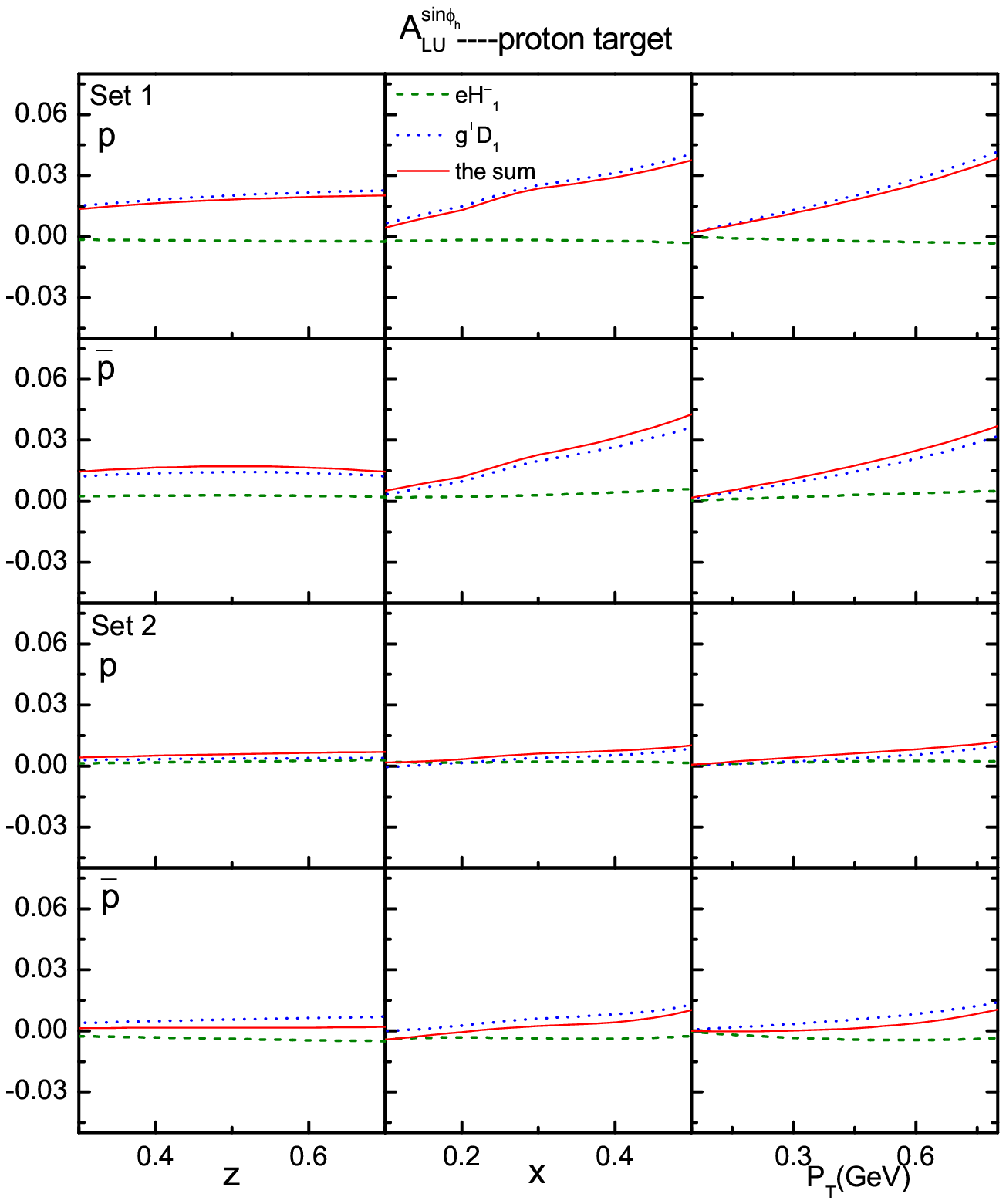}
  \caption{Predictions on the beam SSAs for charged pions (left panel), charged kaons (central panel) and proton/anti-proton (right panel) in SIDIS at JLab with a 12 GeV electron beam scattered off a proton target. The upper panels show the results calculated from the TMD DFs in Set 1 and the lower panels show the results calculated from the TMD DFs in Set 2. The dashed, dotted and solid curves show the asymmetries from the $e H_1^\perp$ term, the $g^\perp D_1$ term and the sum of the two terms, respectively.}
  \label{clas12a}
\end{figure*}
\begin{figure*}
  \includegraphics[width=0.67\columnwidth]{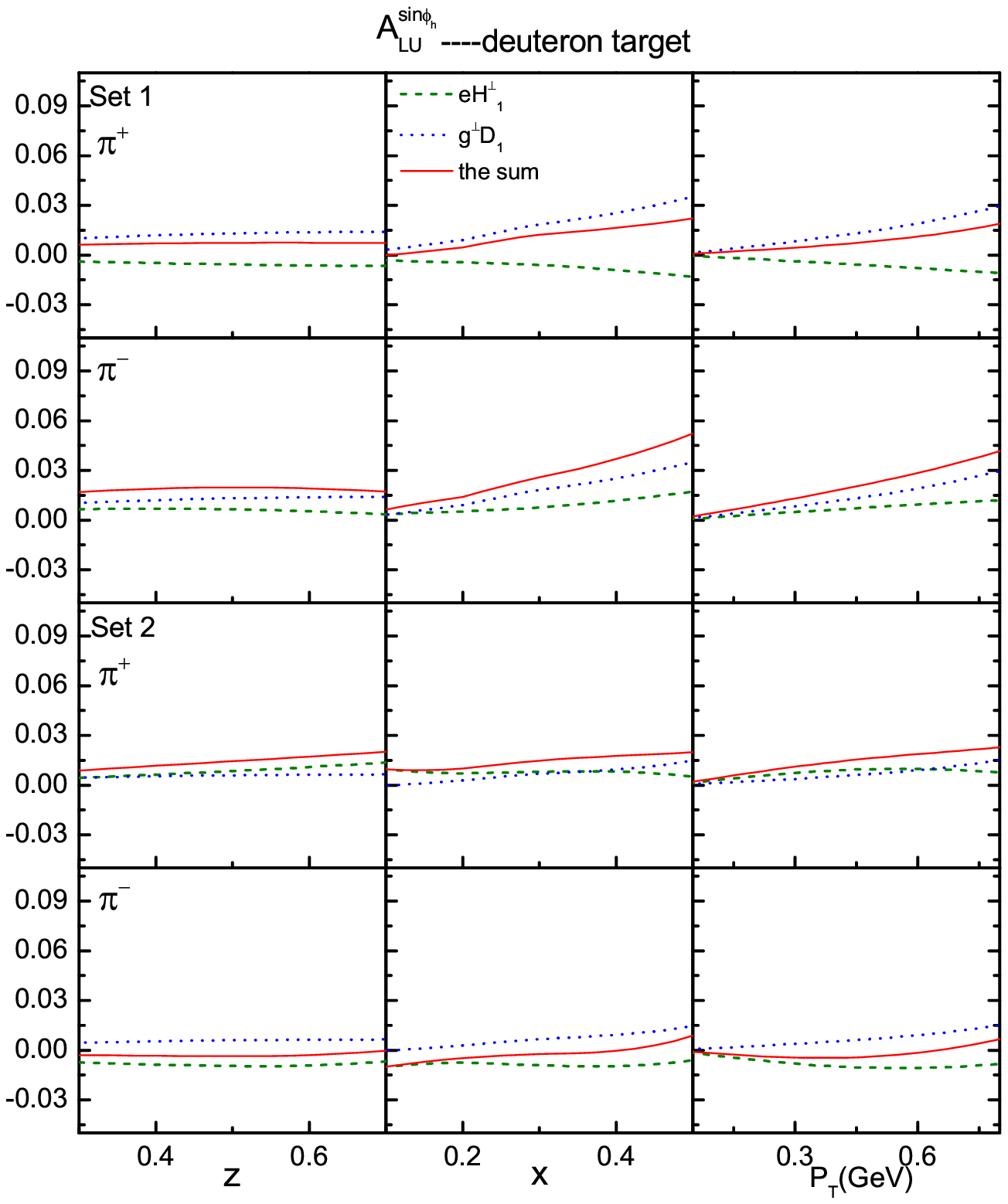}
  \includegraphics[width=0.67\columnwidth]{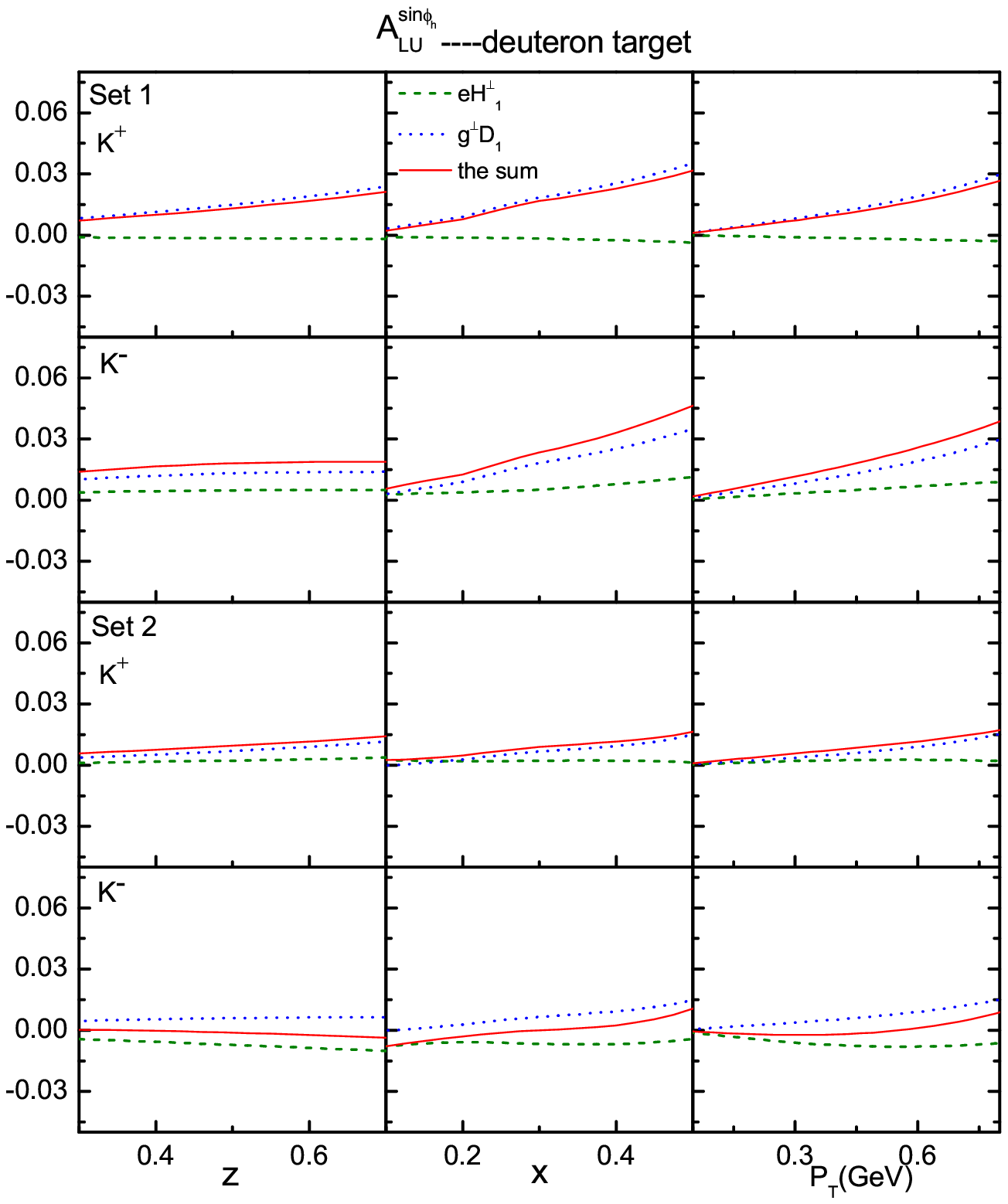}
  \includegraphics[width=0.67\columnwidth]{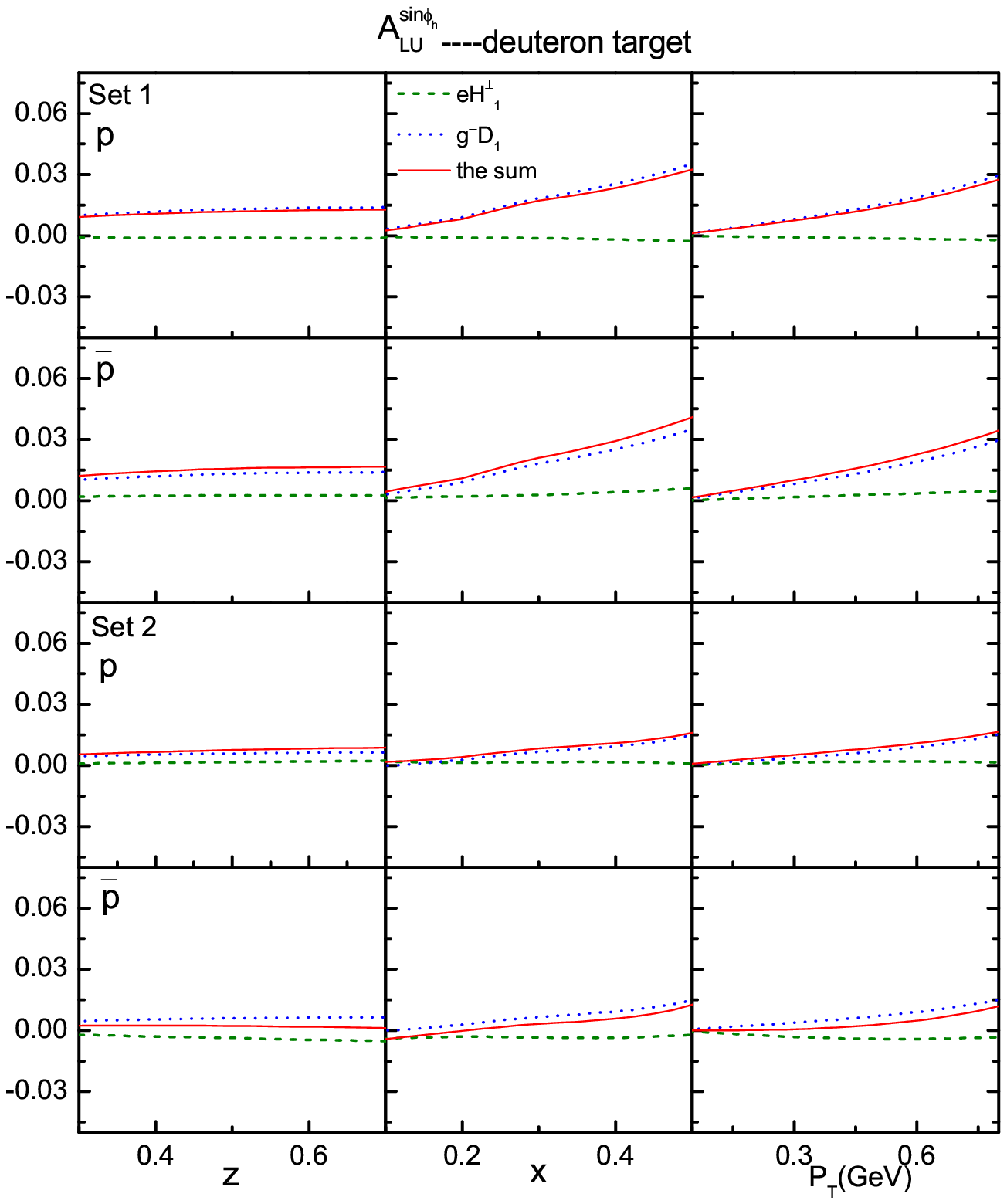}
  \caption{Similar to Fig.~\ref{clas12a}, but with a deuteron target.}
  \label{clas12b}
\end{figure*}
In this subsection, we present our predictions on the beam SSAs for charged hadron production at JLab with a $12 \,\textrm{GeV}$ longitudinally polarized electron beam scattered off nucleon targets, which could be performed in the near future.
We adopt the constraints on $k_T$ given in Eq.~(\ref{constraints}) and apply the following kinematic cuts in the calculation~\cite{Avakian13}:
\begin{align}
&0.1<x<0.6,~~ 0.4<z<0.7,~~ Q^2>1\, \textrm{GeV}^2,\nonumber\\
&P_T>0.05\,\textrm{GeV},~~ W^2>4\,\textrm{GeV}^2.
\end{align}

In Fig.~\ref{clas12a} we plot the beam SSAs for charged hadrons produced in SIDIS by a longitudinally polarized electron beam with 12 GeV scattered off an unpolarized proton target at JLab, as functions of $z$, $x$, and $P_T$.
In our previous work~\cite{wjmao:2012prd}, we already presented the results for $\pi^0$ production at JLab 12 GeV, where we considered the $g^\perp D_1$ term and used the distribution $g^\perp$ calculated in Set 1.
Here we show the beams SSAs for $\pi^+$ and $\pi^-$ in Set 1 and Set 2, in the left panel of Fig.~\ref{clas12a}.
The result for $\pi^+$ production at JLab 12 GeV in Set 1 shows that the asymmetries contributed by two different sources almost cancel, leading to a rather small total asymmetry. In the other cases the pion asymmetries do no vanish. Similarly, we plot the asymmetries for $K^\pm$ and $p/\bar{p}$ in the central and the right panel of Fig.~\ref{clas12a}. We find that the asymmetries for $K^\pm$ and $p/\bar{p}$ in Set 1 are quite sizable, while the the asymmetries for those hadrons in Set 2 are consistent with zero. Therefore, the precise measurements on the beam SSAs for $K^\pm$ and $p/\bar{p}$ production at JLab 12 GeV could be used to distinguish different spectator models. For completeness, in Fig.~\ref{clas12b} we plot the same asymmetries for different charged-hadron production at JLab 12 GeV, but on the deuteron target, in the case that a deuteron target would be available. We find that the size and the sign of the asymmetries on the deuteron target is similar to the case of the proton target.
\section{Conclusion}
\label{conclusion}
In this work, we performed an analysis on the beam SSAs for $\pi^{\pm}$, $K^{\pm}$,  proton and antiproton in SIDIS at the kinematics of HERMES, as well as at the kinematics of JLab 12 GeV. We considered the case that the nucleon target is a proton or a deuteron.
In our calculation we employed the contributions from the $e H_1^\perp$ term and the $g^\perp D_1$ term, and we used two sets of TMD DFs calculated from two different spectator models.
We compared the theoretical curves with the preliminary data recently obtained by the HERMES Collaboration.
We find that for pion production, two sets of TMD DFs lead to
rather different results, also, the roles of the $e H_1^\perp$ term and the
$g^\perp D_1$ term are different in different Sets.
The asymmetries for charged kaons, proton and antiproton are small in both sets and are consistent with the preliminary HERMES data.
For the deuteron target, we find that the role of the $e H_1^\perp$ term is small compared to the $g^\perp D_1$ term.
Therefore, the contribution to beam SSAs related to the $g^\perp D_1$ term
could be studied without a significant background from the mechanism related to the $e H_1^\perp$ term.
Finally, the analysis on the beam asymmetries of charged hadron production at JLab indicates that the precise measurement on the beam SSAs of $K^\pm$ and $p/\bar{p}$ production, which can be performed at JLab with a 12 GeV electron beam in the near future, could be used to distinguish different spectator models and shed light on the mechanism of the beam SSAs in terms of TMD DFs.

\section*{Acknowledgements}
We are grateful to the HERMES Collaboration, in particular to Vitaly Zagrebelnyy for providing the preliminary HERMES data and useful discussion.
This work is partially supported by National Natural Science
Foundation of China (Grant Nos.~11005018 and~11120101004),
by SRF for ROCS from SEM, and by the Fundamental Research
Funds for the Central Universities (Grant No.~2242012R3007).
W. Mao is supported by the Scientific Research Foundation of Graduate School of SEU (Grant No.~YBJJ1336) and by the Research and Innovation Project (Grant No.~CXZZ13$\_$0079) for College Postgraduate of Jiangsu Province.

\end{document}